\documentclass[twocolumn,showpacs,aps,prl,superscriptaddress]{revtex4}

\usepackage{graphicx}
\usepackage{dcolumn}
\usepackage{amsmath}
\usepackage{epsfig}

\input babarsym

\def\Btag {\ensuremath{B_{\mbox{tag}}}\xspace}
\def\Brec  {\ensuremath{B_\textrm{rec}}\xspace}
\def\Btag  {\ensuremath{B_\textrm{tag}}\xspace}

\def\KZ      {\ensuremath{K^0}\xspace}
\def\KKKs      {\ensuremath{K^+ K^- \KS}\xspace}

\def\phiKs     {\ensuremath{\phi \KS}\xspace}

\def\spks{\ensuremath{S_{\phi \KS}}\xspace}
\def\cpks{\ensuremath{C_{\phi \KS}}\xspace}
\def\spkl{\ensuremath{S_{\phi \KL}}\xspace}
\def\cpkl{\ensuremath{C_{\phi \KL}}\xspace}
\def\spk{\ensuremath{S_{\phi K}}\xspace}
\def\cpk{\ensuremath{C_{\phi K}}\xspace}
\def\skk{\ensuremath{S_{KKK}}\xspace}
\def\ckk{\ensuremath{C_{KKK}}\xspace}

\def\finalscbnolabel{\ensuremath{+0.50\pm 0.25^{+0.07}_{-0.04}} \xspace}
\def\finalscb{\ensuremath{+0.50\pm 0.25\, (\mbox{\small stat})^{+0.07}_{-0.04} (\mbox{\small syst})} \xspace}

\def\finalccbnolabel{\ensuremath{0.00\pm 0.23\pm 0.05} \xspace}

\def\finalskknolabel{\ensuremath{-0.42\pm 0.17\pm 0.03} \xspace}

\def\finalckknolabel{\ensuremath{+0.10\pm 0.14\pm 0.04} \xspace}
\def\finalstbkk{\ensuremath{+0.55\pm 0.22\, (\mbox{\small stat}) \pm 0.04\, (\mbox{\small syst}) \pm 0.11\, (\mbox{\small $CP$})} \xspace}
\def\cat{k}
\long\def\inst#1{\par\nobreak\kern 4pt\nobreak
    {\it #1}\par\vskip 10pt plus 3pt minus 3pt}

\newcommand{\BABARPubYear}    {04}
\newcommand{\BABARPubNumber}  {051}

\newcommand{\SLACPubNumber}   {11022}
\newcommand{\LANLNumber}      {0502019}

\def\figurebox#1#2#3{%
    \def\arg{#3}%
    \ifx\arg\empty
    {\hfill\vbox{\hsize#2\hrule\hbox to #2{\vrule\hfill\vbox to #1{\hsize#2\vfill}\vrule}\hrule}\hfill}%
    \else
    {\hfill\epsfbox{#3}\hfill}%
    \fi}

\begin{document}

\preprint{\babar-PUB-\BABARPubYear/\BABARPubNumber} 
\preprint{SLAC-PUB-\SLACPubNumber} 

\begin{flushleft}
hep-ex/\LANLNumber\\[10mm]
\end{flushleft}

\title{
\Large \bf Measurement of {\boldmath \CP} Asymmetries in {\boldmath $ \Bz\to\phi\KZ$}\ and
{\boldmath $ \Bz\to\KKKs$}\ Decays
}

%
\author{B.~Aubert}
\author{R.~Barate}
\author{D.~Boutigny}
\author{F.~Couderc}
\author{Y.~Karyotakis}
\author{J.~P.~Lees}
\author{V.~Poireau}
\author{V.~Tisserand}
\author{A.~Zghiche}
\affiliation{Laboratoire de Physique des Particules, F-74941 Annecy-le-Vieux, France }
\author{E.~Grauges-Pous}
\affiliation{IFAE, Universitat Autonoma de Barcelona, E-08193 Bellaterra, Barcelona, Spain }
% ...
\collaboration{The \babar\ Collaboration}
\noaffiliation

\date{\today}

\begin{abstract}

We measure the time-dependent \CP asymmetry parameters in $\Bz\to\Kp\Km\KZ$
based on a data sample of approximately 227 million 
$B$-meson pairs recorded at the $\Upsilon(4S)$ resonance with the \babar\ detector 
at the \pep2\ $B$-meson Factory at SLAC.
We reconstruct two-body \Bz decays to $\phi(1020)\KS$ and $\phi(1020)\KL$, and
the three-body decay $\KKKs$ with $\phi(1020)\KS$ excluded.
For the $\Bz\to\phi\KZ$ decays, we measure 
$\sin2\beta_\textrm{eff}(\phi\KZ) = \finalscb$.
The $\Bz\to\Kp\Km\KS$ decays are dominated by $K^+K^-$ $S$-wave, as determined from
an angular analysis; we measure
$\sin2\beta_\textrm{eff}(\KKKs) = \finalstbkk$,
where the last error is due to the uncertainty in the fraction of \CP-even contributions
to the decay amplitude. We find no evidence for direct \CP violation.
\end{abstract}


\maketitle

\label{sec:Introduction}
In the Standard Model (SM) of particle physics, the decays $\Bz \to K^+K^-\KZ$~\cite{charge} 
are dominated by $b\rightarrow s\bar{s}s$ gluonic penguin amplitudes, but can also be
affected by amplitudes that are suppressed by elements of the Cabibbo--Kobayashi--Maskawa
(CKM) quark mixing matrix~\cite{ckm}.  These CKM-suppressed amplitudes
cannot be precisely known in a model-independent way~\cite{sPenguin}, but are in general expected
to be small~\cite{phases}. Let $2\beta_\textrm{eff}$ be the \CP-violating
phase difference between decays with and without mixing, and 
$\beta = \arg{(-V_{cd} V^*_{cb}/V_{td}V^*_{tb})}$ where $V_{ij}$ are elements of
the CKM quark mixing matrix. The difference $|\beta-\beta_\textrm{eff}|$ is
expected to be nearly zero, with theoretical uncertainties of a few degrees for
$\Bz\to\phi\KZ$~\cite{phi}. Larger uncertainties exist for $\Bz\to\KKKs$ with
$\Bz\to\phi\KS$ decays excluded, due in part to an extra CKM-suppressed tree
amplitude contribution~\cite{phases}.

Since additional decay diagrams with non-SM particles and interactions
introducing new
\CP-violating phases may contribute to $\beta_\textrm{eff}$, measurements 
of $\sin2\beta_\textrm{eff}$ in these channels and their comparisons with the SM 
expectation are sensitive probes for physics beyond the SM~\cite{phases}.
The value of $\sin2\beta$ has been measured in $\Bz\to J/\psi \KS$~\cite{jpsinew,bellejspinew}
with an average of $0.742\pm0.037$.
The \babar\ and Belle collaborations have measured $\sin2\beta_{\rm eff}$
in $\phi\KZ$ ($+0.47\pm0.34^{+0.08}_{-0.06}$ with 114 million
\BB\ pairs~\cite{phiprl} and $-0.96\pm0.50^{+0.09}_{-0.11}$ with 152 million
\BB pairs($\phi\KS$ only)~\cite{Abe:2003yt}, respectively),
and in $\KKKs$ excluding $\phi\KS$
($+0.57 \pm 0.26 \pm 0.04 ^{+0.17}_{-0.00}$ with 122 million \BB
pairs~\cite{Aubert:2004ta} and
$+0.51 \pm 0.26 \pm 0.05^{+0.18}_{-0.00}$ with 152 million \BB pairs~\cite{Abe:2003yt},
respectively).

At $B$ factories, the neutral $B$ mesons are exclusively produced in pairs.
We select events for which one $B$ ($\Brec$) is reconstructed as
$\Bz\to K^+K^-\KZ$ and the other ($\Btag$) is partially reconstructed
as either $\Bz$ or $\Bzb$. We define $\deltat =  t_\textrm{rec} - t_\textrm{tag}$
to be the difference between the proper decay times of the \B\ mesons.
The decay rate $\mbox{f}_+(\mbox{f}_-)$ for the final state $f$
when the \Btag decays as a \Bz(\Bzb) is given by
\begin{eqnarray}
{\mbox{f}}_\pm(\, \deltat)& = &{\frac{{\mbox{e}}^{{- \left| \deltat 
\right|}/\tau_{\Bz} }}{4\tau_{\Bz}}}  \, [
\ 1 \hbox to 0cm{}
\pm \nonumber\\&&
S_f \sin{( \deltamd  \deltat )}   
\mp 
\,C_f  \cos{( \deltamd  \deltat) }   ], 
\label{eq::timedist}
\end{eqnarray}
where $\tau_{\Bz}$ is the \Bz\ lifetime and \deltamd\ is the
\Bz--\Bzb mixing frequency. The parameter $S_f$ is non-zero
if there is \CP violation in the interference between decays
with and without mixing, while a non-zero value for $C_f$
would entail direct \CP violation.
In the limit where the CKM-suppressed amplitudes do
not contribute, the SM predicts no direct \CP violation
($C_f=0$) since the dominant decay amplitudes have the same \CP-violating
phase, and that $S_f=-\eta_f\times\sin2\beta_\textrm{eff}$.
For $\Bz\to\phi\KS$ decays, the effective eigenvalue $\eta_f=-1$,
for $\Bz\to\phi\KL$ $\eta_f=+1$.
For $\Bz\to\KKKs$ decays, $\eta_f = 2 f_\textrm{even}-1$,
where $f_\textrm{even}$ is the fraction of \CP-even
contributions to the $\Bz\to\KKKs$ amplitude. Then the
value of $\eta_f$ depends on the angular
momentum of the $K^+K^-$ system: it is -1 for relative
$P$-wave and +1 for $S$-wave.

In this paper, we present a measurement of $\sin2\beta_{\rm eff}$ with
almost twice the number of events as for the previous
\babar\ results~\cite{phiprl,Aubert:2004ta}.
We reconstruct \Bz candidates in two independent modes, $\phi\KZ$ (with
the $\KZ$ either a $\KL$ or a $\KS$) and $\KKKs$
(with the $\phi$ mass region excluded). $\KS$'s are detected via their
$\pi^+\pi^-$ decay only.
We extract the \CP\ asymmetry parameters using extended maximum-likelihood fits.
Using an angular moment analysis~\cite{Costa:1980ji}, we extract the 
$K^+K^-$ $P$-wave fractions in the data. These fractions are used to check
the assumption that $\eta_f=-1$ for $\phi\KS$ and +1 for $\phi\KL$ by
bounding the $S$-wave contamination in the $\phi$ mass region, and to
measure $\eta_f$ for $\KKKs$.

\label{sec:event_selection}
This analysis is based on 
227 million \BB\ pairs collected with the \babar\ detector~\cite{Aubert:2001tu} at
the \pep2\ asymmetric-energy \epem\ storage rings at SLAC, operating at the
$\Upsilon(4S)$ resonance (center-of-mass (c.m.) energy $\sqrt{s}=10.58\gev$).
In Ref.~\onlinecite{Aubert:2001tu} we describe 
the silicon vertex tracker (SVT) and drift chamber (DCH) used for track and vertex
reconstruction, the electromagnetic calorimeter (EMC) and instrumented
flux return (IFR) used for $\KL$ reconstruction, and the detector of
internally reflected Cherenkov light (DIRC),
which, together with the EMC, the IFR, and the ionization
$dE/dx$ from the SVT and DCH, is used for particle identification.

The $\Bz$-candidate reconstruction and selection is similar to that described in
Refs.~\onlinecite{phiprl} and~\onlinecite{Aubert:2004ta}.
We consider a $K^+K^-$ pair to be a $\phi$ candidate if its invariant mass is within
$15\mevcc$ (about three times the apparent width in the $K^+K^-$ invariant mass spectrum)
of the central $\phi$ mass value~\cite{pdg2004}.
For a given $\BzBzb$ meson pair, we obtain \deltat\ from the measured distance
between the fully reconstructed \Brec meson decay point and the \Btag
decay point along the beam direction, and the known boost
of the $\Upsilon(4S)$ system ($\beta\gamma$=0.56).
A multivariate tagging algorithm determines the flavor of the
\Btag meson~\cite{jpsinew} and classifies it in one
of seven mutually exclusive tagging categories.

We use two kinematic variables to discriminate between signal $B$ 
decays and combinatorial background.
The energy difference between the measured $\epem$ c.m.
energy of the $B$~candidate and $\sqrt{s}/2$ is $\DeltaE$.
Its distribution peaks at zero for signal,
with a width of about $20\mev$ for $\phi\KS$ and $\KKKs$. The width is only
about $3\mev$ for $\phi\KL$, because for this mode we constrain
the $\Bz$ candidate's mass to the nominal value~\cite{pdg2004}.
The beam-energy-substituted mass, $\mes$, is used for candidates without a $\KL$.
It is defined as
$\mes\equiv\sqrt{(s/2+{\mathbf {p}}_i\cdot{\mathbf{p}}_B)^2/E_i^2-{\mathbf {p}}_B^2},$
where the $B$ momentum ${\mathbf {p}}_B$ and the four-momentum of the 
initial state ($E_i$, ${\mathbf {p}}_i$) are defined in the laboratory frame. 
It peaks at the \Bz mass for signal, with a width of about $3\mev$.
For $\phi\KS$ candidates, we require $|\DeltaE|<100$~\mev and $\mes>5.21~\gevcc$;
for $\phi\KL$ candidates, we require $|\DeltaE|<80$~\mev;
and for $\KKKs$ candidates, we require $|\DeltaE|<200$~\mev and $\mes>5.2~\gevcc$.

The dominant background is continuum $e^+e^-\to q\bar{q}$ ($q = u,d,s,c$) events;
these tend to be jet-like in the $\epem$ c.m. frame, while \B decays tend
to be spherical.
To enhance discrimination between signal and continuum, we 
use Fisher discriminants ($\cal F$) to combine four event-shape-related
variables~\cite{Aubert:2004ta,phiprl}.
The other background originates from $B$ decays. 
For the $\phi \KZ$ final state,
opposite-\CP contributions from the $K^+K^-K^0$ final state 
($K^+K^-$ $S$-wave) are estimated from data with a moment analysis~\cite{Costa:1980ji}
(see below) to be less than 6.6\% at $95\%$ confidence level. 
The mode $\phi \KL$ has additional background. Its dominant $\CP$
contamination is from the mode $\phi\Kstarz\to\phi\KL\piz$, for which
we expect approximately eight events in the region $|\DeltaE|<10\mev$.
In the final likelihood fit we explicitly parameterize
backgrounds from \B decays both with and without charm.

For the $\KKKs$ mode,
we apply invariant mass cuts to suppress background from $B$ decays
that proceed through a $b \to c$ transition, namely those containing
$\Dz$, $\jpsi$, $\chi_{c0}$, or $\psi(2S)$ decaying
into $\Kp\Km$, or $\Dp$ or $\Ds$
decaying into $\Kp\KS$. Finally, to suppress \B decays into final states
with pions, we require the rate for a charged pion to be misidentified
as a kaon to be less than 2\%.

A total of $4\textrm{,}300$ , $8\textrm{,}238$, and $27\textrm{,}368$
events have a candidate that passes the $\phi\KS$, $\phi\KL$, 
or $\KKKs$ selection criteria, respectively.
From simulation, we find the final selection
efficiencies for signal to be $40\%$, $20\%$ and $26\%$, respectively.

\label{sec:cp_asymmetry}
We extract the $\KKKs$, $\phiKs$, and $\phi \KL$ event yields and \CP
parameters with two extended maximum-likelihood fits. One 
is to the $\KKKs$ candidates; the other is to both
the $\phi\KS$ and $\phi\KL$ candidates, with the
assumption that $\cpks = \cpkl$ and $\spks = -\spkl$.
We verified the fit procedure for the $\phi\KZ$ mode with 
samples of $\phi K^+$ and $\jpsi K^0$ events.
We found for the former a null asymmetry as expected,
and for the latter results that are consistent with
previous measurements~\cite{jpsinew}.
We verified the fit procedure for the $\KKKs$ mode with 
a sample of $\KS\KS K^+$ events, for which we found a
null asymmetry as expected.

The likelihood function used in each extended maximum-likelihood fit
to its $N_\cat$ candidates tagged in category $k$ is
\begin{equation}
\label{eq:pdfsum}
{\cal L}_k = e^{-N^{\prime}_\cat}\!\prod_{i=1}^{N_\cat}
		\bigg\{ N_{S} 
	\epsilon_\cat {\cal P}_{i,\cat}^{S}
	+ N_{C,\cat} {\cal P}_{i,\cat}^{C} 
 	+ \sum_{j=1}^{n_B} N_{B,j} \epsilon_{j,\cat}{\cal P}^{\B}_{i,j, \cat}
	\bigg\}
\end{equation}
where $N^{\prime}_\cat$ is the sum of the signal, continuum,
and $n_B$ $B$-background yields tagged in category $k$;
$N_S$ is the number of $\phiKs$, $\phi\KL$, or $\KKKs$ signal
events; $\epsilon_\cat$ is the 
fraction of signal events tagged in category $k$; $N_{C,\cat}$ 
is the number of continuum background events tagged in category $k$;
$N_{B,j}$ is the number of $B$-background events of class $j$;
and $\epsilon_{j,\cat}$ is the fraction of
$B$-background events of class $j$ tagged in category $k$.
Each $B$-background class comprises similar $B$ decays.
The \B-background event yields are fixed parameters and are zero
for the $\phiKs$ sample. The total likelihood 
${\cal L}$ is the product of the likelihoods for each tagging category.

The probability density functions (PDFs)  ${\cal P}_{\cat}^{S}$,  
${\cal P}_{\cat}^{C}$, and ${\cal P}^{\B}_{j, \cat}$, for signal,
continuum background, and $B$-background class $j$, respectively,
are the products of the PDFs of the discriminating variables.
The signal PDF is thus given 
for the $\KKKs$ sample by
${\mathcal P}(\mes) \cdot {\mathcal P}(\DeltaE) \cdot {\mathcal P}({\cal F}) \cdot {\mathcal P}(\deltat;\sigma_{\deltat})$,
for the $\phi \KS$ sample by
$\mathcal{P}(m_{ES})\cdot\mathcal{P}(\Delta E)
\cdot\mathcal{P}(\mathcal{F})
\cdot\mathcal{P}(m_{KK})
\cdot\mathcal{P}(\cos\theta_H)
\cdot\mathcal{P}(\Delta t;\sigma_{\Delta t})$,
and for the $\phi \KL$ sample by
$\mathcal{P}(\Delta E)
\cdot\mathcal{P}(\mathcal{F})
\cdot\mathcal{P}(m_{KK})
\cdot\mathcal{P}(\cos\theta_H)
\cdot\mathcal{P}(\Delta t;\sigma_{\Delta t})$,
where $\theta_H$ is the angle between the $K^+$ candidate
and the parent $\Brec$ flight direction in the $K^+K^-$ rest frame.
The quantity $\sigma_{\Delta t}$ is the
uncertainty in the measurement of $\Delta t$ for a given event.
The time-dependent \CP parameters defined in Eq.~(\ref{eq::timedist}),
diluted by the effects of mistagging and the~$\deltat$ resolution,
are contained in ${\cal P}_\cat^{S}(\deltat, \sigma_{\Delta t})$. 
As in our $\jpsi\KS$ analysis~\cite{jpsinew}, the $\deltat$-resolution function
for signal and $B$-background events is a sum of three Gaussian
distributions, which have two distinct means as well as three distinct widths.
The widths are the error of the measured $\Delta t$ scaled by three independent
factors.

In the fits to data, we leave unconstrained the parameters describing the
\CP\ asymmetry, the $\Delta t$-resolution functions,
the tagging characteristics, and the event yields.
We also leave unconstrained the means of the signal $\mes$ and $\DeltaE$ Gaussian PDFs,
the widths of the signal $\DeltaE$ PDFs, the mean of the signal $m_{KK}$ PDF
(which is parameterized by a relativistic $P$-wave Breit-Wigner function),
and all parameters of the $\KKKs$ candidates' signal PDF for ${\cal F}$.
We take from simulation any other parameters of the $\mes$, $\DeltaE$, ${\cal F}$,
$\cos\theta_H$, and $m_{KK}$  PDFs for signal and $B$ background.
The parameters describing the signal and $B$-background $\Delta t$-resolution
function are determined by a simultaneous fit to an independent sample of
reconstructed $\Bz$ decays to flavor eigenstates, with more than
$100\textrm{,}000$ events~\cite{jpsinew}.
We use the world-averaged values for $\tau_{\Bz}$ and $\deltamd$~\cite{pdg2004}.
The fits to the $\phi\KZ$ and $\KKKs$ candidates have a total of 35 and 34 free
parameters, respectively.

\label{sec:cp_content}

We use an angular moment analysis based on the
$\cos\theta_H$ distribution to extract the $\KKKs$ \CP\ content, and
also to bound the $S$-wave contamination in the $\phi$ mass region.
In this approach, we expand the decay distribution for a given $\Kp\Km$
invariant mass in terms of moments $\left < P_{\ell} \right  >$ of
conveniently normalized Legendre polynomials $P_\ell(\cos\theta_H)$:
\begin{eqnarray}
	|{\cal A}(m_{KK})|^2 	&=& \sum\limits_{\ell} \left < P_\ell \right > \cdot P_\ell(\cos\theta_H),
\label{eq::Pl}
\end{eqnarray}
where ${\cal A}(m_{KK})$ is the mass-dependent decay amplitude.
We normalize $P_\ell(\cos\theta_H)$ such that the integral of 
$P_\ell(\cos\theta_H)^2$ over $\cos\theta_H$ from $-1$ to $1$
equals unity.
We extract the moments by summing over all events:
\begin{equation}
\left < P_\ell \right > ~=~ \sum_j P_\ell(\cos\theta_{H,j}) ~{\cal W}_j / \varepsilon_j,
\label{eq::mom_sPlot}
\end{equation}
where ${\mathcal W}_j$ is the weight for event $j$ to belong to the signal decay
and is calculated by the sPlot technique of Ref.~\onlinecite{Pivk:2004ty}.
The efficiency $\varepsilon_j$ is evaluated from a large MC sample in
bins of $m_{KK}$ and $\cos\theta_H$.
Limiting ourselves to the two lowest partial waves, 
we can write the total decay amplitude in terms of 
the $S$-wave (\CP-even) and the $P$-wave (\CP-odd) amplitudes,
\begin{eqnarray}
	{\cal A}(m_{KK}) &\approx& {\cal A}_S(m_{KK}) P_0(\cos\theta_H) ~+ \nonumber\\&&
                             e^{i\phi_p} {\cal A}_P(m_{KK}) P_1(\cos\theta_H),
\label{eq::SP-waves}
\end{eqnarray}
where $\phi_p$ is the relative phase between the real partial-wave
amplitudes ${\cal A}_S(m_{KK})$ and ${\cal A}_P(m_{KK})$.
If we compare Eq.~(\ref{eq::SP-waves}) to Eq.~(\ref{eq::Pl}),
we can relate the moments (of order $\ell\le2$) to the wave intensities and
thus to the total fraction of \CP-even events, $f_{\rm even}$, as
\begin{eqnarray}
	f_{\rm even} &=&  \frac{{\cal A}_S(m_{KK})^2}{{\cal A}_S(m_{KK})^2 + {\cal A}_P(m_{KK})^2}\nonumber\\
               &=& 1 - \sqrt{\frac{5}{4}} \frac{ \left< P_2 \right >}{\left< P_0 \right > }, 
\label{eq::feven_moments}
\end{eqnarray}
where ${\cal A}_S(m_{KK})^2$ and ${\cal A}_P(m_{KK})^2$ are the $S$- and $P$-wave intensities,
respectively. In the normalization, the total number of signal events is
$\sqrt{2}\left < P_{0} \right  >$.

\label{sec:Systematics}

\begin{table}[ht]
\caption{Systematic uncertainties on the \CP\ parameters.}
\begin{tabular}{lcc|cc}
\hline \hline
Source    & \spk & \cpk & \skk & \ckk \\
\hline
Detector effects  & $\pm0.02$ & $\pm0.02$ & $\pm0.02$ & $\pm0.01$ \\ 
DCSD                  & $\pm0.01$ & $\pm0.03$ & $\pm0.00$ & $\pm0.03$ \\ 
Fit bias              & $\pm0.01$ & $\pm0.01$ & $\pm0.02$ & $\pm0.01$ \\ 
$\Bz$-$\Bzb$ tagging  & $\pm0.01$ & $\pm0.02$ & $\pm0.00$ & $\pm0.01$ \\
$S$-wave contamination& $+0.06$   & $\pm0.02$ & -         & -         \\
Other    & $\pm0.03$ & $\pm0.02$ & $\pm0.01$ & $\pm0.01$ \\
\hline
Total                    &  $^{+0.07}_{-0.04}$ & $\pm0.05$ & $\pm0.03$           & $\pm0.04$ \\
\hline \hline
\end{tabular}
\label{tb::systCP}
\end{table}

Systematic errors on the \CP-asymmetry parameters are listed in Table
\ref{tb::systCP}. We account for uncertainties in the \deltat
resolution, the beam-spot position, and the detector alignment. We
also estimate errors due to the effect of doubly CKM-suppressed decays
(DCSD) of the \Btag~\cite{Long:2003wq}.
The uncertainty due to possible biases in the fit
procedure is conservative and includes
effects on the \CP\ parameters of correlations among the fit
variables, which have been determined with full-detector MC simulations.
Uncertainties in the $\Bz$-$\Bzb$ tagging efficiency in both signal
and background are also included.
Finally, we account for errors due to the \CP content of the background,
uncertainties in the PDF parameterization, and the uncertainties of
$\tau_{\Bz}$ and $\deltamd$~\cite{pdg2004}.
For each mode we add the individual contributions in quadrature to obtain
the total systematic uncertainty.

We also consider the systematic error due to the \CP-even fraction of the $\KKKs$ mode.
We do not find evidence for the existence of higher moments $\left < P_\ell \right >$,
$\ell=3\dots 6$, which could arise from intermediate $D$-wave decays into $\Kp\Km$ or decays
proceeding through an isospin-$1$
resonance into $\Kpm\KS$.  Nevertheless, we estimate a systematic error from the $D$-wave by
examining the $\left < P_2 \right >$ moment in the $K^+K^-$ mass region (1.1--1.7) \gevcc,
corresponding to the $f_2(1270)$, $a_2(1320)^0$, and $f_2'(1525)$ resonances,
and assuming that $\left < P_2 \right >$ arises only from $D$-wave
and $S$-$D$ interference.  Since the moment itself is consistent with zero, we assign a systematic
error of 4\% based on the $\left < P_2 \right >$ error. 
We account for the possible presence of $a_0(980)^+$,
$a_0(1450)^+$, and $a_2(1320)^+$ in the $\Kpm\KS$ subsystem ($4.6\%$).
We also estimate a bias due to the modeling of the efficiency from MC events~(2.5\%). 
We find the total systematic error on $f_\textrm{even}$ to be $\pm0.06$. This leads
to a systematic error on $\sin2\beta_\textrm{eff}$ of $\pm0.11$.

\begin{figure}[!htb]
\begin{center}
\epsfig{file=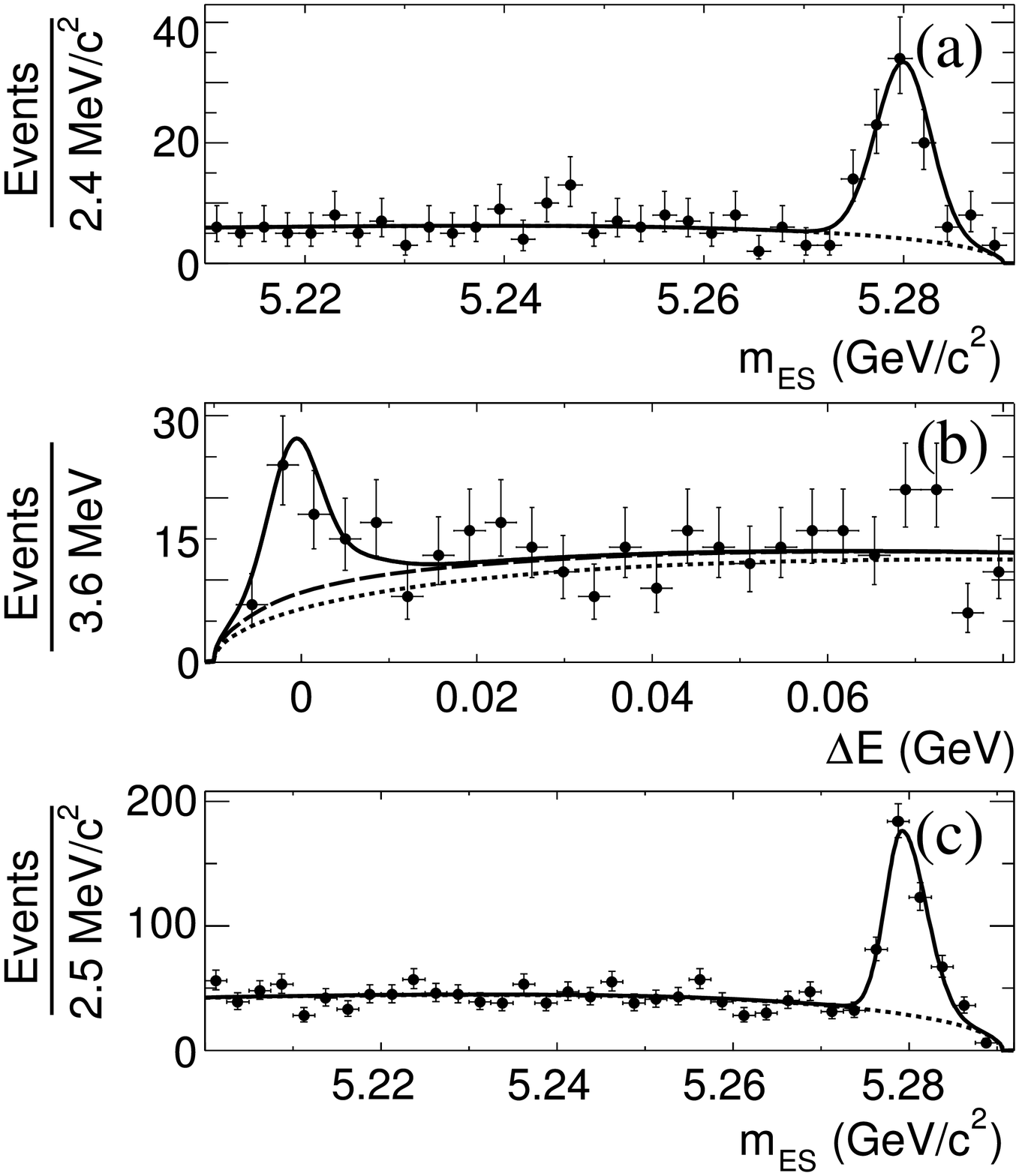,width=8.0cm}
\vspace{-0.40cm}
\caption{
Distributions of (a) \mes for $\phiKs$ candidates,
(b) \DeltaE for $\phi\KL$ candidates, and (c) \mes for $\KKKs$
candidates excluding $\phi\KS$, together with the results from the final
extended maximum-likelihood
fits after applying a requirement on the ratio of signal likelihood to 
signal-plus-background likelihood (computed without the displayed variable)
to reduce the background. The requirement is chosen to roughly maximize
$N_S^2/(N_S+N_C)$ where $N_C$ is the total number of continuum events, and
is applied only for the purpose of making these plots. The
curves are projections from the
likelihood fits for total yield (solid lines), continuum background
(short dashed lines), and total background (long dashes in (b) only).
The efficiency of the likelihood-ratio cut is
(a) $79\%$ for signal and $5\%$ for background,
(b) $35\%$ for signal, $16\%$ for \B-background, and $3\%$ for continuum background, and
(c) $77\%$ for signal and $5\%$ for background.
}
\label{fg::projection_plots}
\end{center}
\vspace{-0.20in}
\end{figure}

\begin{figure}[!hpt]
\begin{center}
\vskip 1cm
\epsfig{file=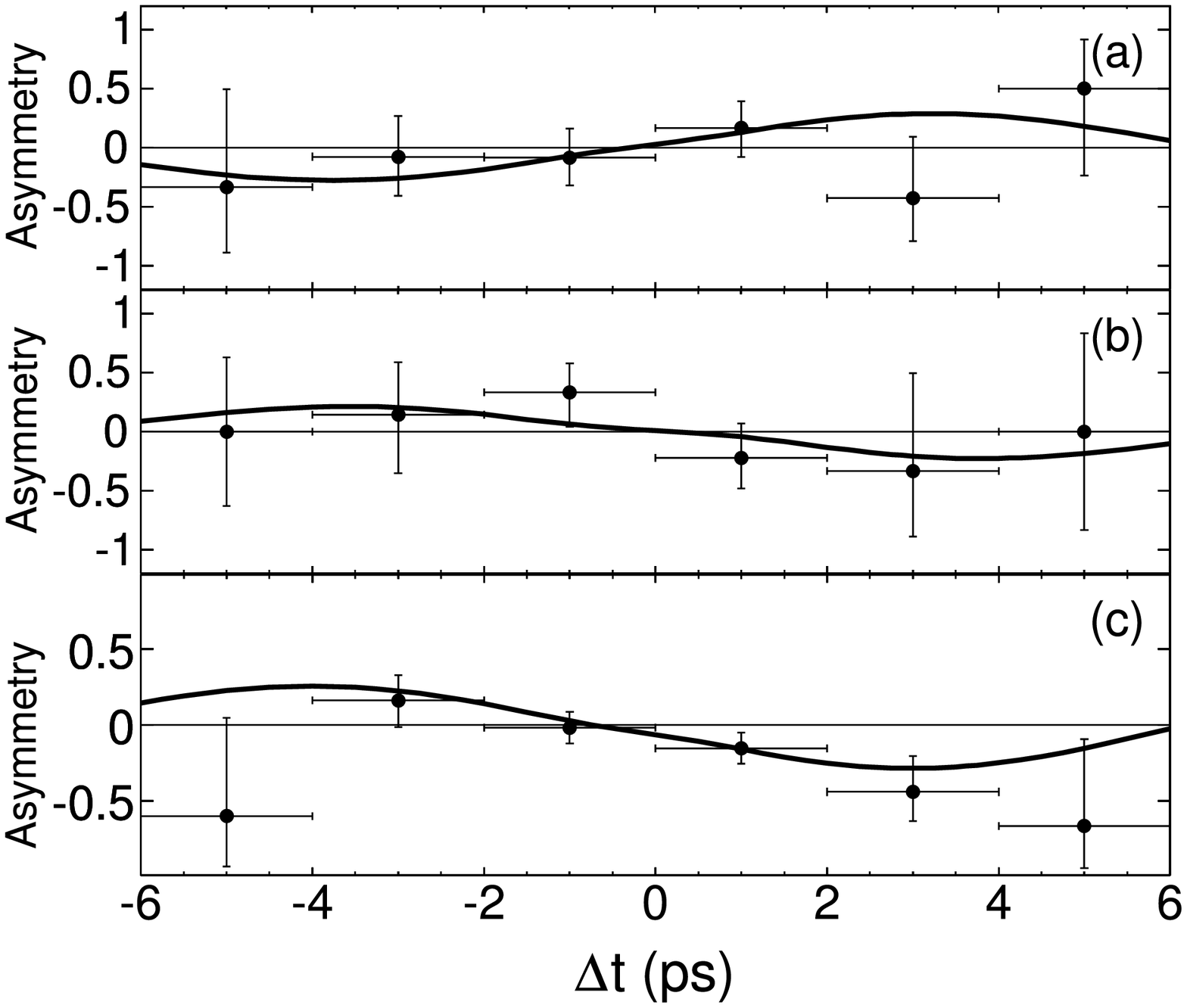, width=8.7cm} 
\vspace{-0.40cm}
\caption{
The  time-dependent asymmetry distributions
for (a) $\phi\KS$, (b) $\phi\KL$, and (c) $\KKKs$ with no $\phi\KS$ decays.
The asymmetry is defined as 
$A_{\Bz/\Bzb}=  (N_{\Bz} - N_{\Bzb})/(N_{\Bz} + N_{\Bzb})$,
where $N_{\Bz}$($N_{\Bzb}$) is the number of \Btag\ mesons identified
as a $\Bz$($\Bzb$) for a given measured value of $\Delta t$.
The signal-to-background ratio is enhanced with a cut on the likelihood
ratio as in Fig.~\ref{fg::projection_plots}.
\label{fg::dt}
}
\end{center}
\vspace{-0.20in}
\end{figure}

\label{sec:results}
Table~\ref{tb::results} shows the measured \CP parameters and yields from the final 
extended maximum-likelihood fits.  Note that when fitting $\sin2\beta_\textrm{eff}$
for $\KKKs$, we constrain $\ckk$ to zero. All yields are consistent with our previously
measured branching fractions~\cite{Aubert:2003hz,Aubert:2004ta}.
Figure~\ref{fg::projection_plots} shows the signal-enhanced distributions of
\mes for $\phi\KS$ and $\KKKs$ events
and of \DeltaE  for $\phi\KL$ events, together with the result from the final extended
maximum-likelihood fits.
Figure~\ref{fg::dt} shows the time-dependent asymmetry distributions.
As a cross check, we also fit $\phi\KS$ and $\phi\KL$ separately.
Our fit to only $\phi\KS$ events gives $S=0.29\pm0.31$ and $C=-0.07\pm0.27$.
Our fit to only $\phi\KL$ events gives $S=1.05\pm0.51$ and $C=0.31\pm0.49$.

\begin{table}[th]
\caption{ \CP-asymmetry parameters and yields from the final extended maximum-likelihood fits,
as well as the fraction of \CP-even contributions to the amplitude, $f_\textrm{even}$, which
is assumed to be zero for $\phi\KS$ and and unity $\phi\KL$.
The first errors are statistical, and the second are systematic;
the third error on $\sin2\beta_\textrm{eff}$ for $\KKKs$ is due to
the uncertainty
in the \CP content. The values of $S$ and $C$ are fit simultaneously 
for the $\phi\KS$ and $\phi\KL$ candidates; the sign of $S$ for $\phi\KS$ is shown.
When finding $\sin2\beta_\textrm{eff}$ for $\KKKs$, we constrain $\ckk$ to 0.
}
\begin{tabular}{lccc}
\hline\hline
                          &\multicolumn{2}{c}{$\phi\KZ$}        &$\KKKs$                           \\
                          &$\phi\KS$          &$\phi\KL$        &      (no $\phi\KS$)              \\\hline
$\sin2\beta_\textrm{eff}$ &\multicolumn{2}{c}{\finalscbnolabel} &$+0.55 \pm 0.22 \pm 0.04 \pm 0.11$ \\\hline
$f_\textrm{even}$         &$0$                &$1$              &$0.89 \pm 0.08 \pm 0.06$          \\
$S$                       &\multicolumn{2}{c}{\finalscbnolabel} &\finalskknolabel                  \\
$C$                       &\multicolumn{2}{c}{\finalccbnolabel} &\finalckknolabel                  \\
Yield                     &$114\pm12$         &$98\pm18$        &$452\pm28$                        \\
\hline\hline
\end{tabular}
\label{tb::results}
\end{table}

\begin{figure}[!htb]
\begin{center}
\includegraphics[height=10cm]{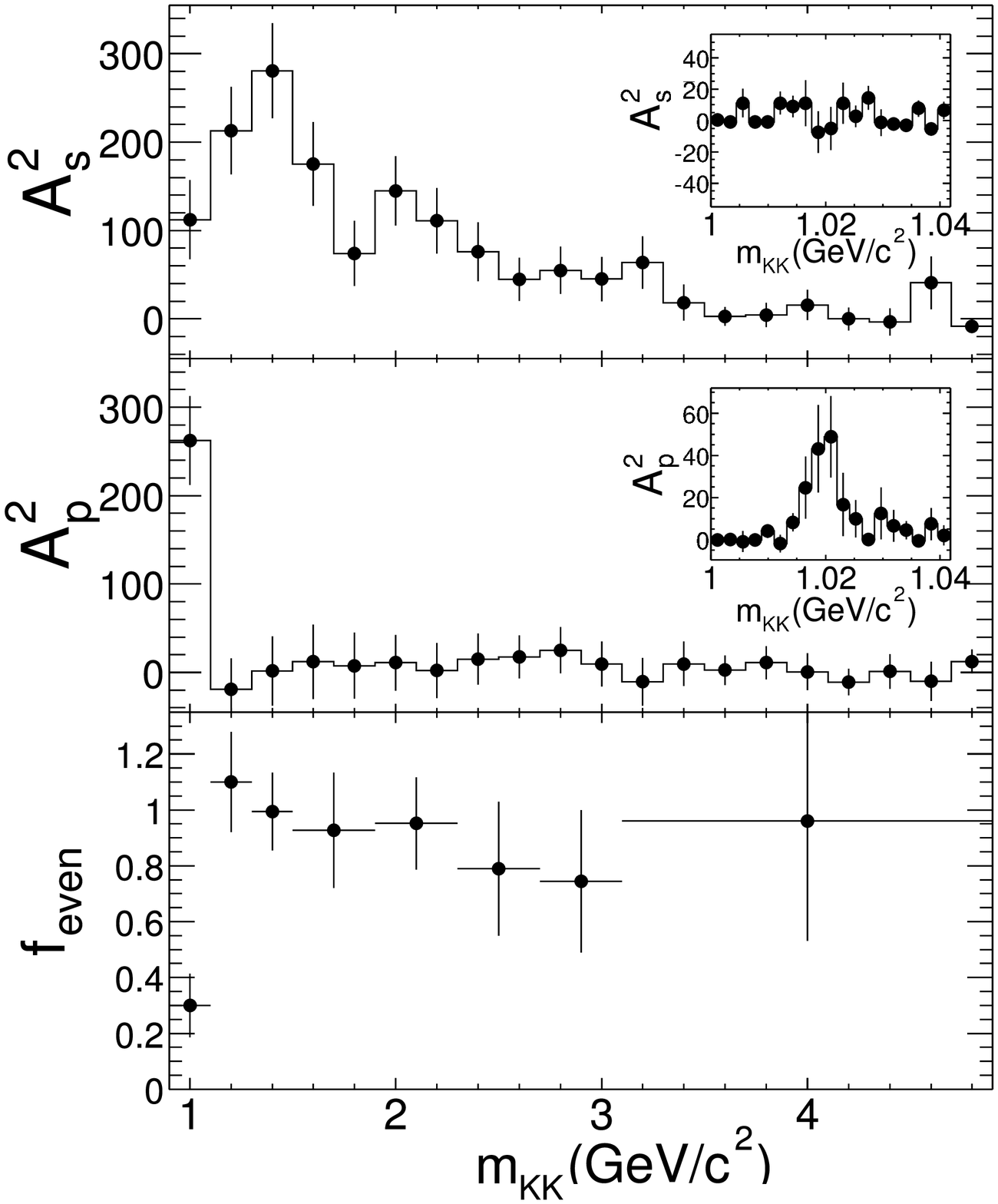}
\vspace{-0.40cm}
\caption{Distributions of $S$- and $P$-wave intensities and
\CP\-even fraction as a function of $\Kp\Km$ invariant mass. 
Notice that the first bin integrates a wider mass range than
the $\phi$ resonance occupies.
Insets show $S$- and $P$-wave intensities in the $\phi$ mass region.
}
\label{fg::waves}
\end{center}
\vspace{-0.20in}
\end{figure}

For the $\KKKs$ final state including the $\phi$ mass region,
the distributions of the $S$- and $P$-wave intensities,
and the \CP-even fraction, as a
function of $\Kp\Km$\ invariant mass, are shown in Fig.~\ref{fg::waves}.
The total fraction of \CP-even events with the $\phi$ mass region excluded
is given in Table~\ref{tb::results}.
We successfully verified our value of $f_\textrm{even}$ with a different
method~\cite{Garmash:2003er} that uses the event rates in 
$B^+ \to K^+ \KS \KS$ and the isospin-related channel $B^0 \to K^+ K^- \KS$.

\label{sec:Summary}
To summarize, in a sample of 227 million \BB meson pairs, we measure the \CP content
and \CP parameters in \Bz-meson decays into $\phi\KZ$, and into $\KKKs$
with the $\phi$ mass region excluded.
We determine the fraction of 
\CP-even and \CP-odd contributions with an angular analysis.
In $\Bz\to\phi\KZ$, our values for $\sin2\beta_{\rm eff}$ and $\cpk$ are
in good agreement with our previously published values~\cite{phiprl},
and the small $S$-wave contamination is treated as a systematic uncertainty.
In $\Bz\to\KKKs$, the $K^+K^-$ system is observed to be dominated by $S$-wave;
this, along with the measured value of $\sin2\beta_{\rm eff}$,
is consistent with previous
measurements based on isospin symmetry~\cite{Abe:2003yt,Aubert:2004ta}.
Both of our $\sin2\beta_{\rm eff}$ values are consistent to within one standard 
deviation with the value of $\sin2\beta$ measured in $\Bz\to c\bar{c} s$
decays~\cite{jpsinew}.

We are grateful for the excellent luminosity and machine conditions
provided by our \pep2\ colleagues, 
and for the substantial dedicated effort from
the computing organizations that support \babar.
The collaborating institutions wish to thank 
SLAC for its support and kind hospitality. 
This work is supported by
DOE
and NSF (USA),
NSERC (Canada),
IHEP (China),
CEA and
CNRS-IN2P3
(France),
BMBF and DFG
(Germany),
INFN (Italy),
FOM (The Netherlands),
NFR (Norway),
MIST (Russia), and
PPARC (United Kingdom). 
Individuals have received support from CONACyT (Mexico), A.~P.~Sloan Foundation, 
Research Corporation,
and Alexander von Humboldt Foundation.


\begin{thebibliography}{99}

\bibitem{charge}
Throughout this paper, charge conjugate reactions are included implicitly.

\bibitem{ckm}
N.~Cabibbo, \jprl {\bfseries 10}, 531 (1963);
M.~Ko\-ba\-ya\-shi and T.~Maskawa, \progtp {\bfseries 49}, 652 (1973).

\bibitem{sPenguin}
D.~London and R.~D.~Peccei,
Phys.\ Lett.\ B {\bf 223}, 257 (1989)%
;
N.~G.~Deshpande and J.~Trampetic,
Phys.\ Rev.\ D {\bf 41}, 895 (1990)%
;
R.~Fleischer,
Z.\ Phys.\ C {\bf 62}, 81 (1994)%
;
N.~G.~Deshpande and X.~G.~He,
Phys.\ Lett.\ B {\bf 336}, 471 (1994)
;
Y.~Grossman, Z.~Ligeti, Y.~Nir and H.~Quinn,
Phys.\ Rev.\ D {\bf 68}, 015004 (2003)
;
M.~Gronau and J.~L.~Rosner,
Phys.\ Lett.\ B {\bf 564}, 90 (2003).

\bibitem{phases}
A.B.~Carter and A.I.~Sanda, \jprd{23}, 1567 (1981);
I.I.~Bigi and A.I.~Sanda, \npb{193}, 85 (1981);
R.~Fleischer and T.~Mannel, Phys.\ Lett.\ B {\bf 511}, 240 (2001);
Y.~Grossman, G.~Isidori and M.P.~Worah, Phys.\ Rev.\ D {\bf 58}, 057504 (1998);
Y.~Grossman, Z.~Ligeti, Y.~Nir and H.~Quinn, Phys.\ Rev.\ D {\bf 68}, 015004 (2003);
Y.~Grossman and M.P.~Worah, \plb{395}, 241 (1997);
R.~Fleischer, Int. J. Mod. Phys. A {\bfseries 12}, 2459 (1997);
D.~London and A.~Soni, \plb{407}, 61 (1997).

\bibitem{phi}
Throughout this paper, $\phi$ refers to the $\phi(1020)$.

\bibitem{jpsinew}
\babar\ Collaboration, B.~Aubert {\itshape et al.},
submitted to Phys.~Rev.~Lett,
hep-ex/0408127.

\bibitem{bellejspinew}
Belle Collaboration, K.~Abe {\it et al.},
submitted to Phys.\ Rev.\ D,
hep-ex/0408111.

\bibitem{phiprl}
\babar\ Collaboration, B.~Aubert {\it et al.},
Phys.\ Rev.\ Lett.\  {\bf 93}, 071801 (2004).

\bibitem{Abe:2003yt}
Belle Collaboration, K.~Abe {\it et al.},
Phys.\ Rev.\ Lett.\  {\bf 91}, 261602 (2003).

\bibitem{Aubert:2004ta}
\babar\ Collaboration, B.~Aubert {\it et al.},
Phys.\ Rev.\ Lett.\  {\bf 93}, 181805 (2004).

\bibitem{Costa:1980ji}
G.~Costa {\it et al.},
Nucl.\ Phys.\ B {\bf 175}, 402 (1980);
S.U.~Chung, \jprd{56}, 7299 (1997).

\bibitem{Aubert:2001tu}
\babar\ Collaboration, B.~Aubert {\it et al.},
\nima{479}, 1 (2002).

\bibitem{pdg2004}
Particle Data Group, S.~Eidelman {\it et~al.}, Phys.~Lett.~B {\bf 592}, 1 (2004).


\bibitem{Pivk:2004ty}
M.~Pivk and F.~R.~Le Diberder,
physics/0402083 (2004).

\bibitem{Long:2003wq}
O.~Long, M.~Baak, R.~N.~Cahn, and D.~Kirkby,
Phys.\ Rev.\ D {\bf 68}, 034010 (2003).

\bibitem{Aubert:2003hz}
\babar\ Collaboration, B.~Aubert {\it et al.},
Phys.\ Rev.\ D {\bf 69}, 011102 (2004).

\bibitem{Garmash:2003er}
Belle Collaboration, A.~Garmash {\it et al.},
Phys.\ Rev.\ D {\bf 69}, 012001 (2004).

\end{thebibliography}
\end{document}